# GeneNetMiner: accurately mining gene regulatory networks from literature

Chabane Tibiche and Edwin Wang[1,2]

1. Lab of Bioinformatics and Systems Biology, National Research Council Canada, Montreal, Canada, H4P 2R2
2. Center for Bioinformatics, McGill University, Montreal, Canada

**Abstract**

Lots of gene regulatory relationships have been reported derived 'small-scale' studies. These relations are buried in literature which is rapidly growing. It is extremely time-consuming and cost-intensive to manually curate gene regulatory relationships by human-reading articles. Therefore, a tool for prioritizing these relations and assisting manual curation is useful for finding gene regulatory relationships which could be used for constructing gene regulatory networks. GeneNetMiner is standalone software which parses the sentences of iHOP (Information Hyperlinked over Protein) and captures regulatory relations. The regulatory relations are either gene-gene regulations or gene-biological processes (i.e., Gene A induces cancer metastasis) relations. Capturing of gene-biological process relations is a unique feature for the tools of this kind. These relations can be used to build up gene regulatory networks for specific biological processes, diseases, or phenotypes. Users are able to search genes and biological processes to find the regulatory relationships between them. Each regulatory relationship has been assigned a confidence score, which indicates the probability of the 'true' relation. Furthermore, it reports the sentence containing the queried terms, which allows users to manually checking whether the relation is true if they wish. GeneNetMiner is able to accurately capture the regulatory relationships between genes from literature. The software is available at http://www.cancer-systemsbiology.org.

**Background**

Gene regulatory networks explicitly represent the molecular regulatory mechanisms and encode causality of genes for biological processes and diseases. Analyzing and modeling of gene networks could lead to discovering the emergence of molecular mechanisms of many diseases such as cancer, further helping in identifying potential therapeutic targets [1]. New biological insights and principles of gene regulations can be uncovered from network analysis and modeling [2-5].

It is still challenging to apply high-throughput technologies for profiling gene regulatory relations. On the other hand, a huge amount of gene regulatory relations and gene-biological process (i.e., Gene A induces cancer progression) relations have been uncovered in a low-throughput manner and documented in literature during the past few decades. It is time- and cost-consuming to collect these relations, even impossible by reading literature. Thus, it is important to develop tools to find the gene relations using computer and context-based data mining of scientific literature. Toward this end, several tools have been developed to perform automatic text mining within the PubMed database, e.g. LitMiner [6], PubGene [7], iHOP [8], EBIMed [9]. These tools use diverse search strategies, for example, automatic gene recognition and color coding of keywords. Among these tools, iHOP is one of the most useful tools for this purpose. iHOP is a database containing the protein/gene relations which are embedded in key sentences extracted from literature. Recently, Gene Interaction Miner (GIM) was developed to extract the gene relations from iHOP in an explicit fashion, i.e. Gene A and Gene B have a relation [10]. With these relations, one could construct gene-gene relation network. However, GIM can't tell whether both genes have regulatory relations. Moreover, GIM output contains many noisy data, because it just based on the co-occurrence of the two genes in one sentence. To overcome these problems, we developed a new tool, GeneNetMiner, to accurately extract gene-gene and gene-phenotype or biological processes (i.e, Gene A induces cancer metastasis) regulatory relations from the contextual information provided by iHOP.

**Implementation**

We looked for regulatory relations between gene pairs or between gene-phenotype pairs, or a pair of a gene and a biological process. MeSH (Medical Subject Headings) keywords from PubMed were used as keywords to represent biological processes and phenotypes. We defined a gene, a phenotype or a biological process as an 'object'. We first obtained the sentences stored in iHOP (http://www.ihop-net.org/). The regulatory relations were extracted by walking through the iHOP sentences. For each pair of consecutive objects in an iHOP sentence, we first looked up the regulatory relation indexes (Supplementary Table 1, new regulatory relation indexes can be added into the list) representing either "activation" or "inhibition" between the objects. By analyzing these iHOP sentences, we found that the average length of the characters between two consecutive objects which contain a regulatory relationship is around 10. Thus, we developed a scoring function to measure whether the regulatory relation is most likely true for two objects. If the number of characters is less than 10 between two consecutive objects, we set the score to 9 for the relation; otherwise we calculate the score as [10 – (length of the characters between two objects /10)].

**Results**

GeneNetMiner inputs are either gene names, NCBI gene IDs or MeSH keywords representing cell phenotypes (i.e., diseases) or biological processes. GeneNetMiner outputs contain the regulatory relationships such as activation or inhibition between gene pairs or gene-phenotype/biological processes pairs. Furthermore, the outputs also contain the iHOP sentence containing the queried objects. These sentences allow to manually validating the captured relations quickly. The tool also gives a confident score (range from 1 to 9) for each relation. The scale of the score represents the false positive rate of the GeneNetMiner derived relations.

We evaluated the scoring function by querying the iHOP database (as May, 2010) using whole human genome genes. We randomly checked a few hundreds of GeneNetMiner captured relations by manually reading the iHOP sentences which contain the queried objects. We found that the high-score relations assigned by GeneNetMiner contain more 'correct relations'. To get an idea which score is useful in extracting more accurate relations, we extended the manual check to ~8,000 randomly selected relations. We found that when score is 7, 8 or 9, more than 71%, 91% and 84% of the relations are correct, respectively. When the score is below than 7, the accuracy of GeneNetMiner captured relations goes down dramatically to 30%.

**Discussion and conclusions**

GeneNetMiner is designed for accurately extracting regulatory relationships (positive or negative) from iHOP database. GeneNetMiner has a friendly graphic interface. Users can query a set of the genes/phenotypes/biological-processes of interest. The confident score for each GeneNetMiner-derived relation can be used to help in making decisions whether the relation should be dropped. GeneNetMiner has some unique features that most of the tools of this kind do not have: (1) GeneNetMiner outputs also contain the iHOP sentences containing the queried objects. If users wish, they can quickly scan the sentences to confirm the relation. (2) GeneNetMiner is looking for not only gene-gene relationship, but also gene and biological process relationship, which is another unique feature. Gene networks can be constructed from these relations, furthermore, GeneNetMiner allows to constructing gene network for specific biological processes, diseases, or other phenotypes. Thus far, no such tool has been developed.

For comparison to the most closely related tool, GIM. GIM also extracts gene-gene relations from iHOP, but it is only based on the co-occurrence of two genes. GIM derived gene-gene relations do not indicate regulatory relationships and does not link genes to biological processes or phenotypes. Furthermore, by manually checking GIM outputs and the related iHOP sentence, we found that lots of noises have been captured by GIM. The GeneNetMiner software and the user guide can be downloaded from the journal website.

Availability and requirements

GeneNetMiner is a stand alone, Java-based platform. The platform is free for academic use.

Authors contributions

EW and CT conceived and designed the software. CT implemented and tested the software. EW and CT wrote the manuscript. All authors read and approved the final manuscript.

Acknowledgements

The work is supported by NRC Genomics and Health Initiative.